\documentclass[a4paper]{jpconf}
\pdfoutput=1
\usepackage{graphicx}
\usepackage{hyperref}
\usepackage{iopams}
\usepackage{amsmath,amssymb,amsfonts}
\begin{document}

\title{Effect of dispersity of particle length on the electrical conductivity of the two-dimensional systems}

\author{Yuri~Yu~Tarasevich$^1$, Andrei~V~Eserkepov$^1$, Irina~V~Vodolazskaya$^1$, Petr~G~Selin$^1$, Valentina~V~Chirkova$^1$, Nikolai~I~Lebovka$^{2,3}$}
\ead{tarasevich@asu.edu.ru}
\address{$^1$Astrakhan State University, Astrakhan, 414056, Russia}
\address{$^2$Department of Physical Chemistry of Disperse Minerals, F.~D.~Ovcharenko Institute of Biocolloidal Chemistry, NAS of Ukraine, Kiev, Ukraine, 03142}
\address{$^3$Department of Physics, Taras Shevchenko Kiev National University, Kiev, Ukraine, 01033}

\begin{abstract}
 By means of computer simulation, we examined effect of dispersity of filler length on electrical conductivity of two-dimensional (2D) composites with rod-like fillers. Continuous approach has been used. Highly conductive zero-width rod-like particles were deposited uniformly with given anisotropy onto a poorly conductive substrate. Length of particles varied according to the lognormal distribution. Our simulation evidenced that slightly disordered systems may have both the high figure of merit, i.e., both high electrical conductivity and the optical transmission simultaneously, and the high electrical anisotropy.
\end{abstract}

\section{Introduction}\label{sec:intro}
Transparent and flexible electrodes are important components of modern optoelectronic devices (see, e.g.,~\cite{Hecht2011AM} for review). Such the electrodes are transparent, poorly conductive films containing highly conductive particles.  Recently, simulations of electrical conductivity of two-dimensional systems with rod-like fillers of equal length have been performed both in lattice~\cite{Tarasevich2016PRE,Tarasevich2017PhA,Tarasevich2018JPhCS} and in continuous~\cite{Lebovka2018PRE,Tarasevich2018PRE} approaches. By contrast, fillers in real systems of conductive nanocomposites have different lengths follow a lognormal distribution~\cite{Forro2013AIPA}. The goal of the present conference paper is to exam the effect of length dispersity on the electrical conductivity of the two-dimensional system with rod-like fillers.

\section{Methods}\label{sec:methods}
Continuous approach has been used to prepare a 2D sample. Highly conductive zero-width rod-like particles were deposited uniformly with given anisotropy onto a poorly conductive substrate of size $L \times L$ subjected to periodic boundary conditions, i.e., onto a torus. The rods are allowed to intersect each other. The electrical conductivities of the particles and the substrate were $\sigma_p$ and $\sigma_m$, correspondingly. High electrical contrast $\Delta = \sigma_p/\sigma_m$ was assumed. Length of particles, $l$, varied according to the lognormal distribution.
%Probability density function is
%\begin{equation}\label{eq:PDFLN}
%f(l)=\frac{1}{l \sigma_l \sqrt{2\pi }}\exp \left( -\frac{{{\left( \ln l-{\mu_l} \right)}^{2}}}{2\sigma _{l}^{2}} \right).
%\end{equation}
We performed simulations for the fixed value of the mean value of the length $\langle l \rangle = 1$ and different values of the standard deviation, $\mathrm{SD} = 0, 0.1, 0.5, 1.0$.

The simulation were performed for different values of the order parameter defined as
\begin{equation}\label{eq:s}
s={{N}^{-1}}\sum\limits_{i=1}^{N}{\cos }2{{\theta }_{i}},
\end{equation}
where $\theta_i$ is the angle between the axis of the $i$-th rod and the horizontal axis $x$, and $N$ is the total number of rods in the system. In our simulations, values of the order parameter were $s=0, 0.5, 0.8, 1$.

To calculate the electrical conductivity of the film, the torus was unwrapped and transformed into a random resistor network (RRN) by means of a discretization. A square mesh of $m \times m$ was applied to the film. When a cell of the mesh contained any part of a rod it was assumed to be conductive and opaque. Otherwise, it was assumed to be insulating and transparent. The fraction of conductive (opaque) cells is denoted as $p$. Frank--Lobb algorithm~\cite{Frank1988PRB} was utilised to calculated the conductivity of the RRN.

The percolation threshold, $p_c$, was found using Hoshen--Kopelman algorithm~\cite{Hoshen1976PRB}.

The detailed description of our approach can be found in Ref.~\cite{Tarasevich2018JAP}.

To characterize the electrical anisotropy, we used the quantity~\cite{Tarasevich2016PRE,Tarasevich2017PhysA}
\begin{equation}\label{eq:delta}
  \delta = \frac{\log_{10} \left(\sigma_\parallel / \sigma_\perp\right)}{\log_{10} \Delta},
\end{equation}
where $\sigma_\parallel$ and $\sigma_\perp$ are the electrical conductivity along and perpendicular to the alignment of the fillers, respectively.

The figure of merit~\cite{Haacke1976JAP}
\begin{equation}\label{eq:FoM}
  \Phi_\text{TC} = T^{10}/R_s
\end{equation}
was calculated to characterise the films. Here, $T$ is the optical transmission and $R_s$ is the sheet resistance. In our computations, the optical transmission was taken as $T = 1 - p$. The sheet resistance $R_s$ is defined by
\begin{equation}\label{eq:Rs}
  R_s = 1/\sigma t,
\end{equation}
where $\sigma$ is the electrical conductivity and $t$ is the thickness of the film. Since the thickness  of the film was constant in all our simulations, it was assumed to be 1 arbitrary units in calculations of the figure of merit. Hence, the figure of merit was measured in arbitrary units.

To characterise the electrical properties of films at low concentrations of conductive rods, the ``intrinsic conductivity''
\begin{equation}\label{eq:ic}
[\sigma] = \left.\frac{\mathrm{d}\ln\sigma}{\mathrm{d}p}\right|_{p \to 0}
\end{equation}
was calculated. This quantity was introduced~\cite{Douglas1995AdvChPh,Garboczi1996PRE} by means the following virial expansion
\begin{equation}\label{eq:Garboczi1996}
\frac{\sigma}{\sigma_m} =1+[\sigma]p+ \mathrm{O}\left(p^2\right).
\end{equation}

Throughout the text, the error bars in the figures correspond to the standard deviation of the means. When not shown explicitly, they are of the order of the marker size.

\section{Results}\label{sec:results}
\begin{figure}[!htbp]
($a$)\includegraphics[width=0.45\textwidth]{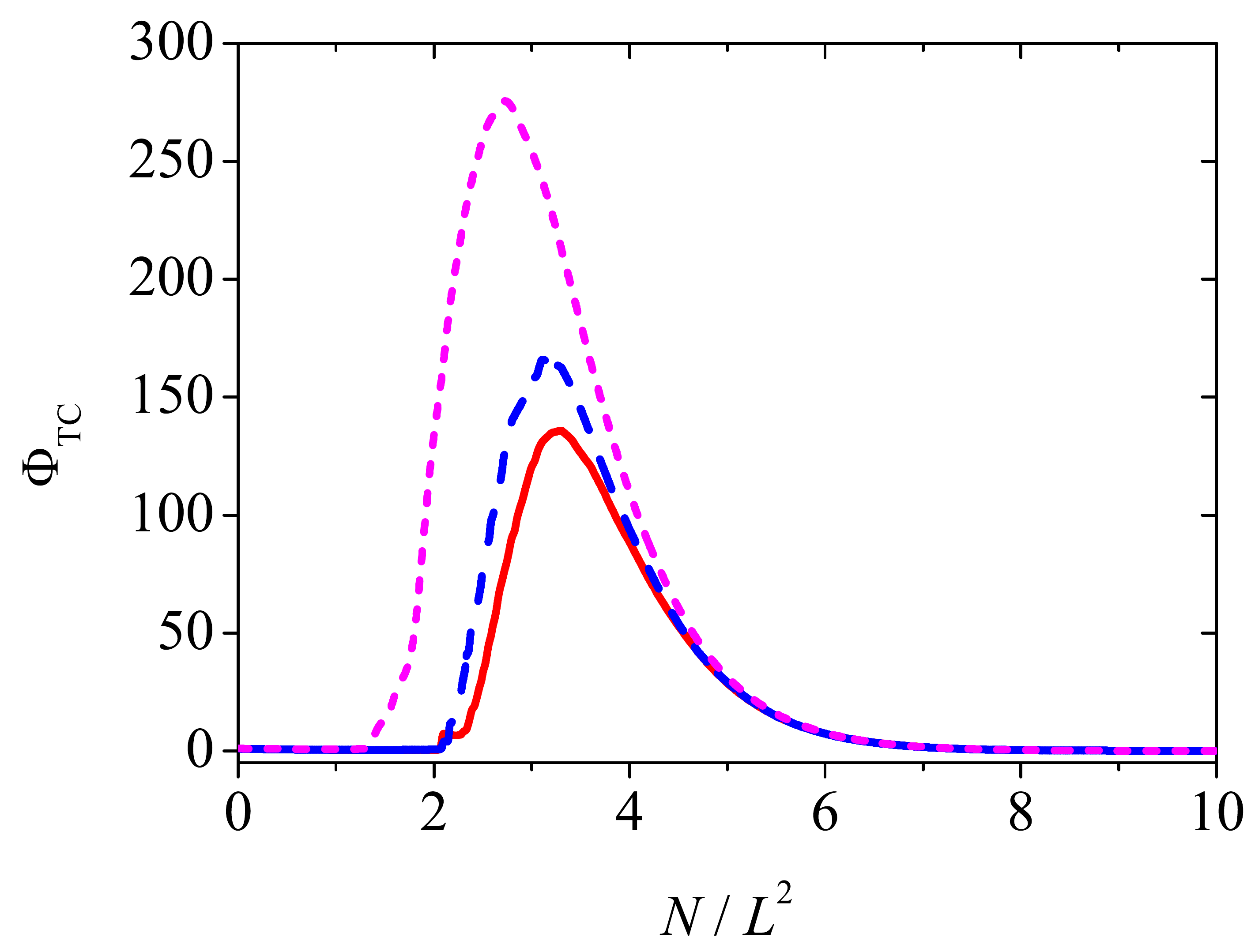}%
($b$)\includegraphics[width=0.45\textwidth]{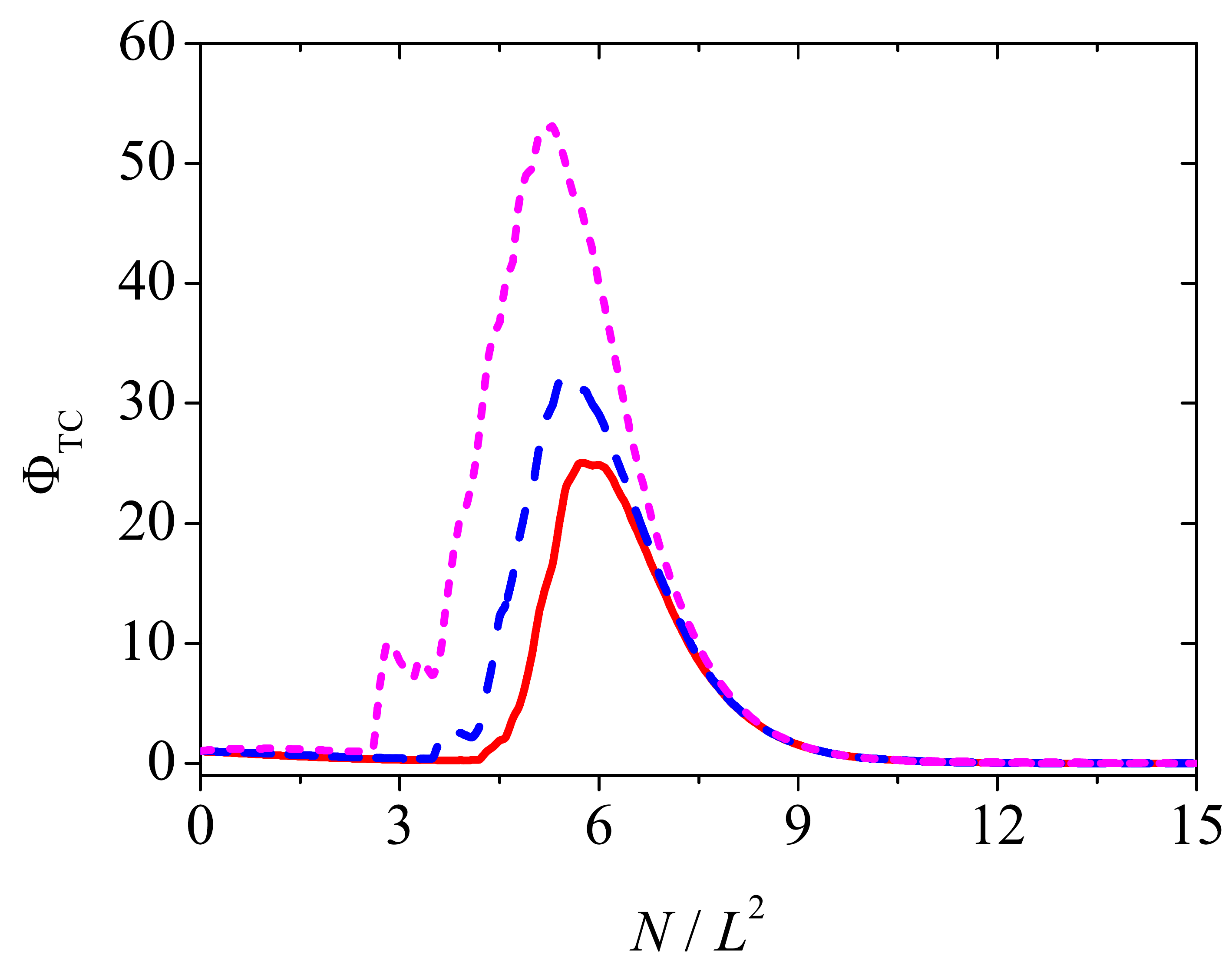}
\caption{\label{fig:FigureOfMerit}Examples of the figure of merit, $\Phi_\text{TC}$, vs number of particles per unite area, $m=256$. \full\ $\mathrm{SD} = 0$, \broken\ $\mathrm{SD} = 0.5$, \dashed\ $\mathrm{SD} = 1$. ($a$)~$s=0$; ($b$)~$s=1$, the electrical conductivity was computed along direction of fillers alignment.}
\end{figure}

\Fref{fig:FigureOfMerit} demonstrates effect of dispersity of filler length on the figure of merit for two values of the order parameter. Both the case of isotropic system ($s=0$) and in the case of strictly anisotropic system, i.e., a system with completely aligned fillers ($s=1$), increase in the  dispersity led to both the increase of the maximal value of the figure of merit and decrease of the number of fillers per area when this maximum occurred. Nevertheless, the figure of merit was significant smaller (about 5 times) for anisotropic system in compare with isotropic case.

\begin{figure}[!htbp]
($a$)\includegraphics[width=0.45\textwidth]{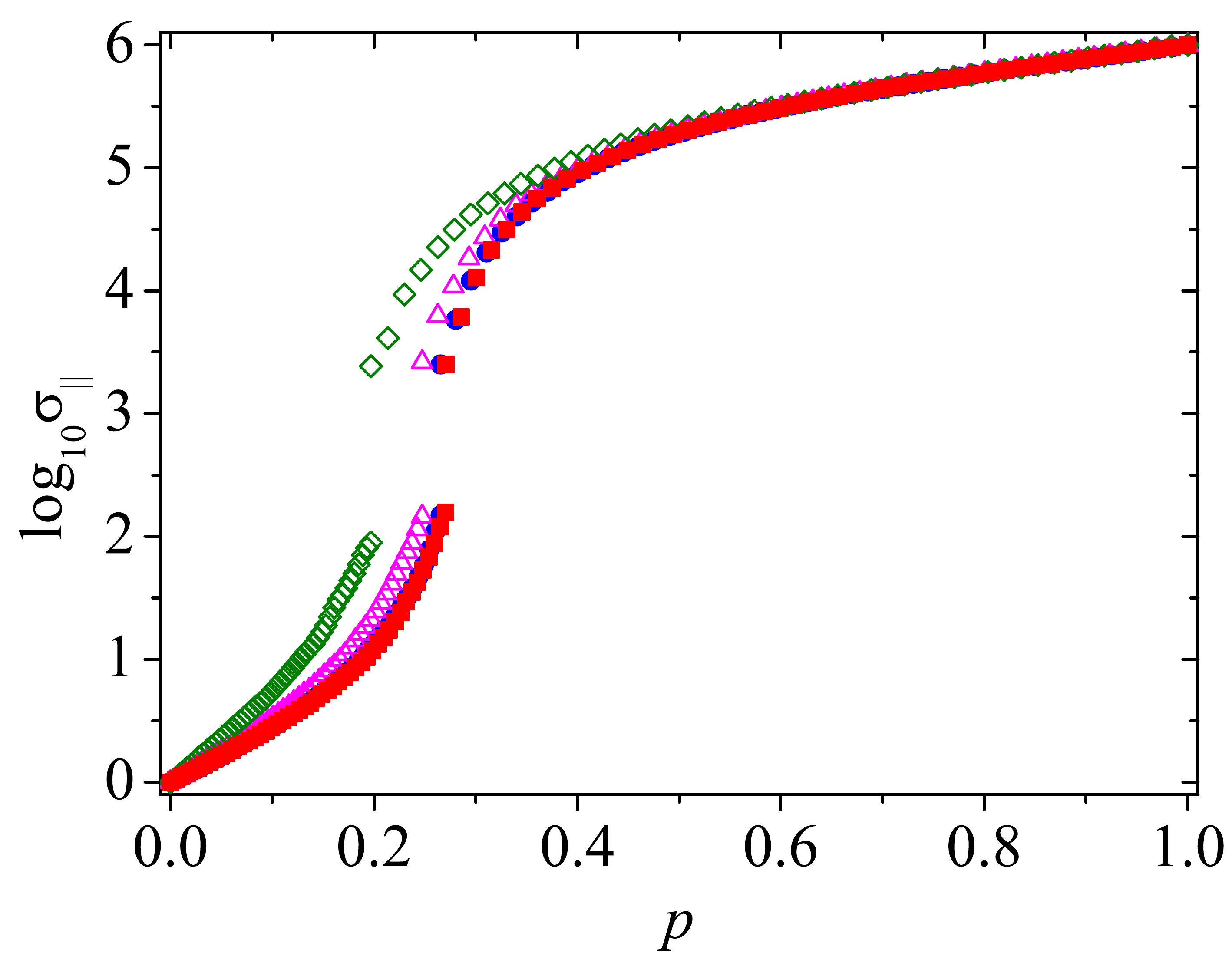}\hfill
($b$)\includegraphics[width=0.45\textwidth]{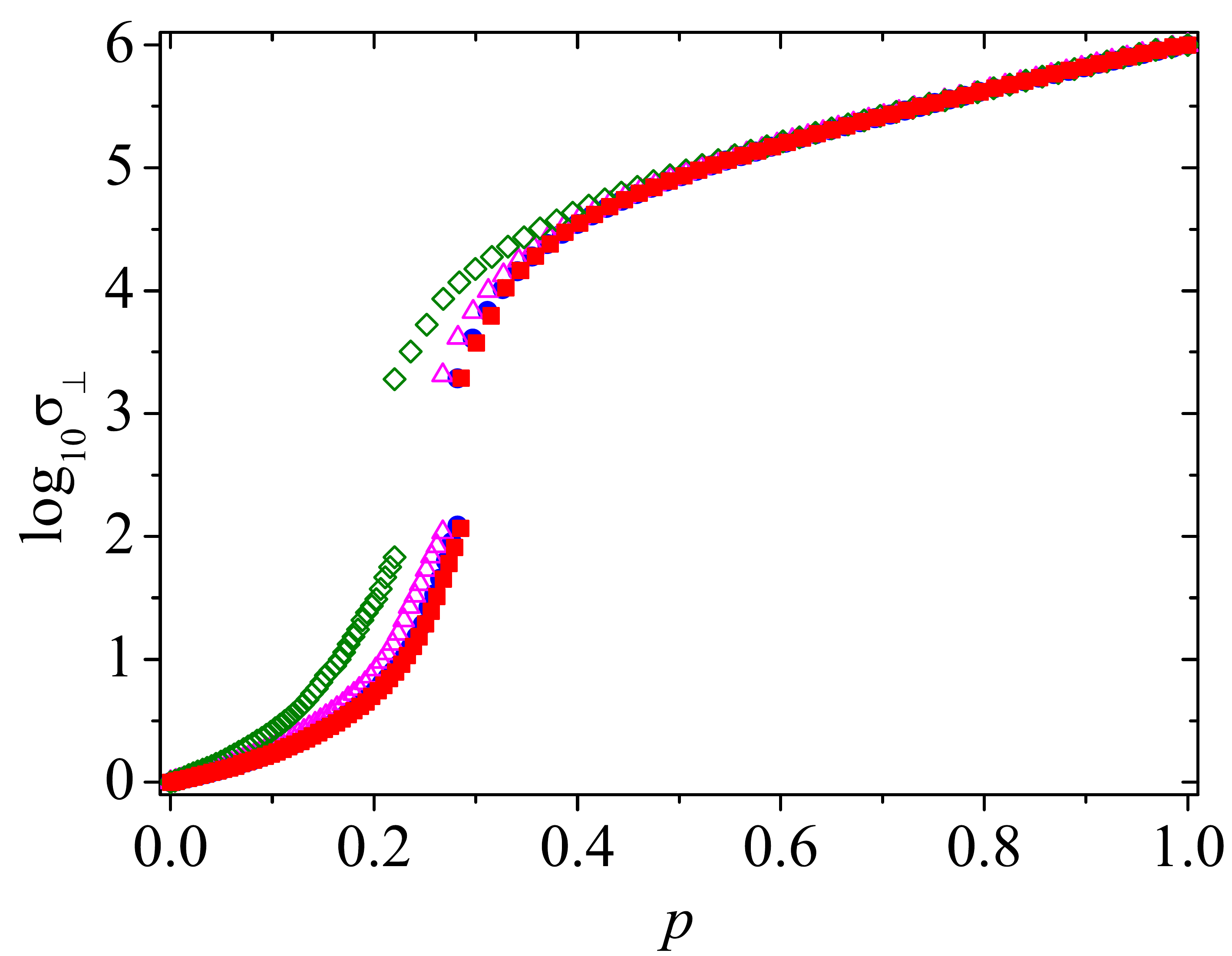}
\caption{\label{fig:conds05m512}Examples of the dependencies of the logarithm of the averaged effective electrical conductivity, $\sigma$, on the concentration, $p$, at  \fullsquare~$\mathrm{SD} = 0$, \fullcircle~$\mathrm{SD} = 0.1$, \opentriangle~$\mathrm{SD} = 0.5$, \opendiamond~$\mathrm{SD} = 1.0$ and $s = 0.5$, $m = 512$, $L = 32$ ($a$)~along the direction of alignment of the rods, ($b$)~in the perpendicular direction.}
\end{figure}

The results presented below are obtained for lattices with $L=32$. \Fref{fig:conds05m512} shows, as an example, the results of a numerical calculation of the dependence of the logarithm of the averaged effective electrical conductivity, $\sigma$,  on the concentration, $p$,  along the direction of alignment of the rods ($a$) and in the perpendicular direction ($b$) at $\mathrm{SD}=0, 0.1, 0.5, 1.0$ and $s=0.5, m=512$. The curves have the shape of a sigmoid; the jump in electrical conductivity indicates an insulator-metal transition. The concentration at which this transition occurs is reduced when the  length dispersity increases in both perpendicular directions at partial ordering of the rods ($s=0.5$). The number of long rods increases when the dispersity increases, which improves the connectivity of the system in both directions.

\begin{figure}[!htbp]
\begin{minipage}[c]{0.45\textwidth}
\includegraphics[width=\textwidth]{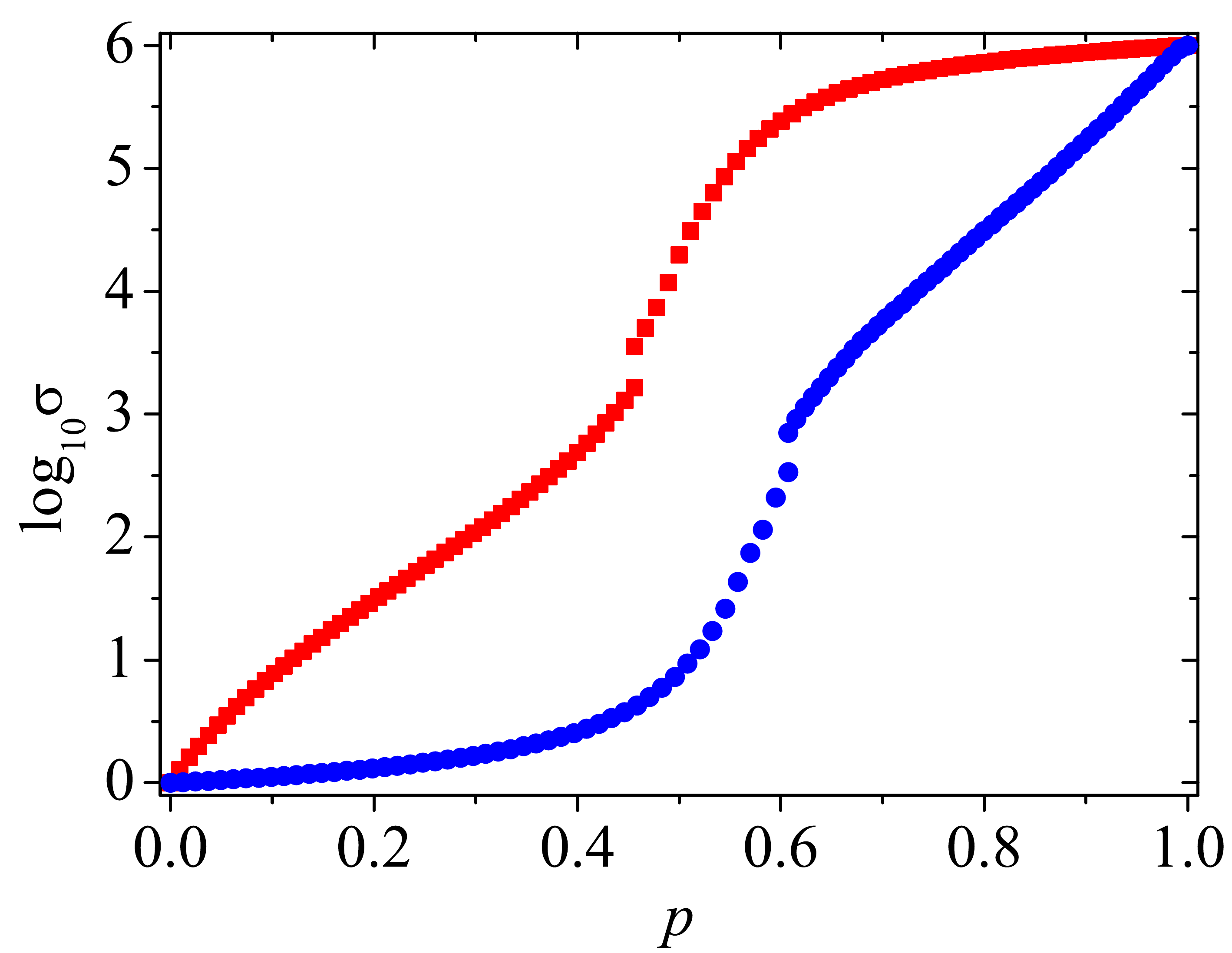}
\end{minipage}
\hfill%
\begin{minipage}[c]{0.45\textwidth}\caption{\label{fig:condSD01s1m1024}Examples of the dependencies of the effective electrical conductivity, $\sigma$, on the concentration, $p$, $\mathrm{SD} = 0.1$, $s = 1$, $m = 1024$, $L = 32$  along the direction of alignment of the rods (\fullsquare) and in the perpendicular direction (\fullcircle).}
\end{minipage}
\end{figure}

For an anisotropic system, the concentrations at which the transition occurs in perpendicular directions must be different, and \fref{fig:condSD01s1m1024} demonstrates this fact. Examples of the dependencies of the averaged effective electrical conductivity, $\sigma$, on the concentration, $p$,  along the direction of alignment of the rods  and in the perpendicular direction are presented in this figure at $\mathrm{SD} = 0.1$, $s = 1$, $m = 1024$. The transition will occur at lower concentrations along the alignment direction of the rods, since the appearance of long rods will affect the connectivity of the system primarily along the alignment direction.

\begin{figure}[!htbp]
($a$)\includegraphics[width=0.45\textwidth]{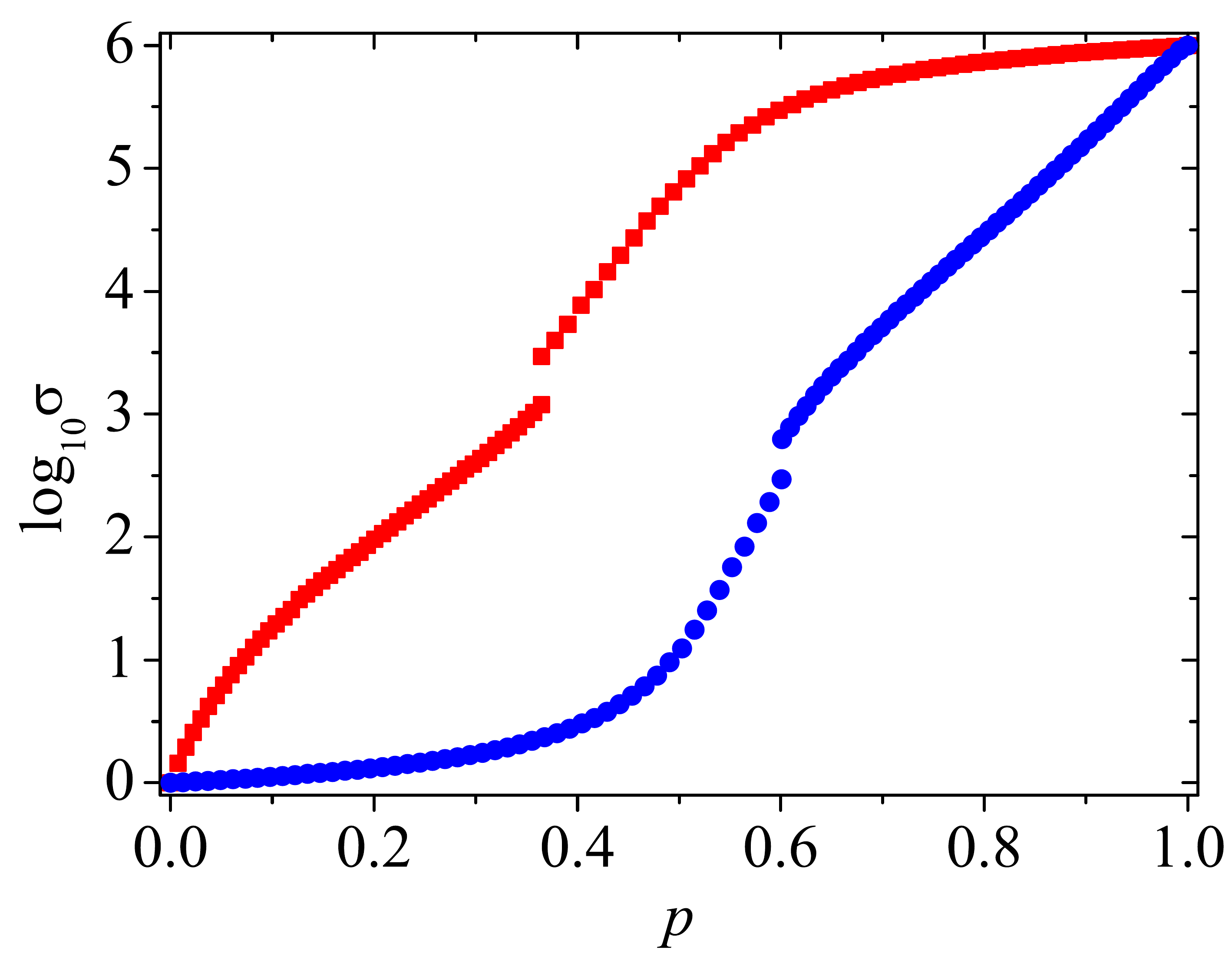}
($b$)\includegraphics[width=0.45\textwidth]{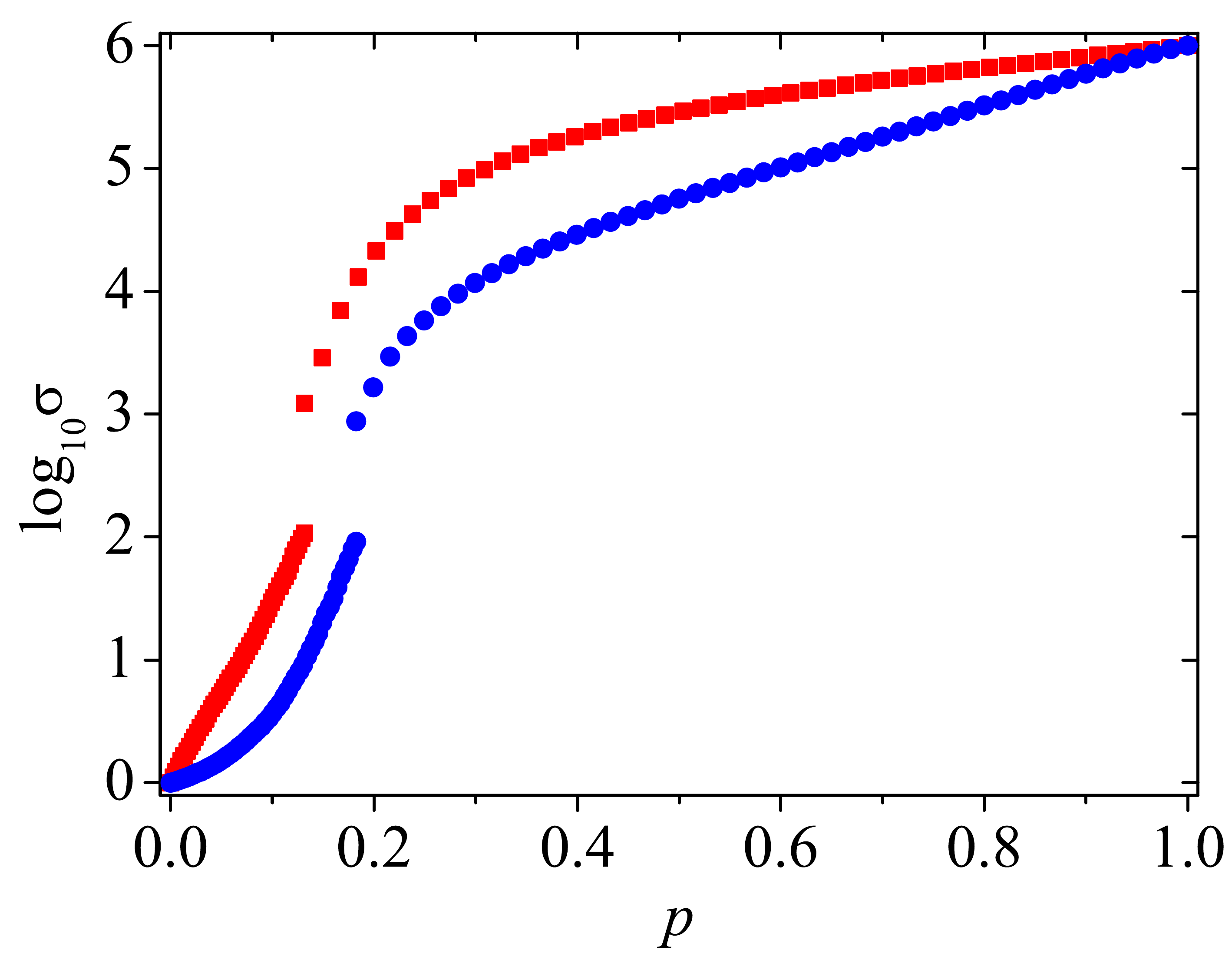}
\caption{\label{fig:condSD1m1024}Examples of the dependencies of the effective electrical conductivity, $\sigma$, on the concentration, $p$, $\mathrm{SD} = 1$, $s = 1$, $m = 1024$, $L = 32$ along the direction of alignment of the rods (\fullsquare) and in the perpendicular direction (\fullcircle): ($a$)~$s = 1$; ($b$)~$s = 0.8$.}
\end{figure}

The more orderly the system is, the stronger this difference becomes (compare \fref{fig:condSD1m1024}(a) for $s=1$  and~\fref{fig:condSD1m1024}(b) for $s=0.8$) at $\mathrm{SD} = 1$, $s = 1$, $m = 1024$.

\begin{figure}[!htbp]
\begin{minipage}{0.45\textwidth}
\includegraphics[width=\textwidth]{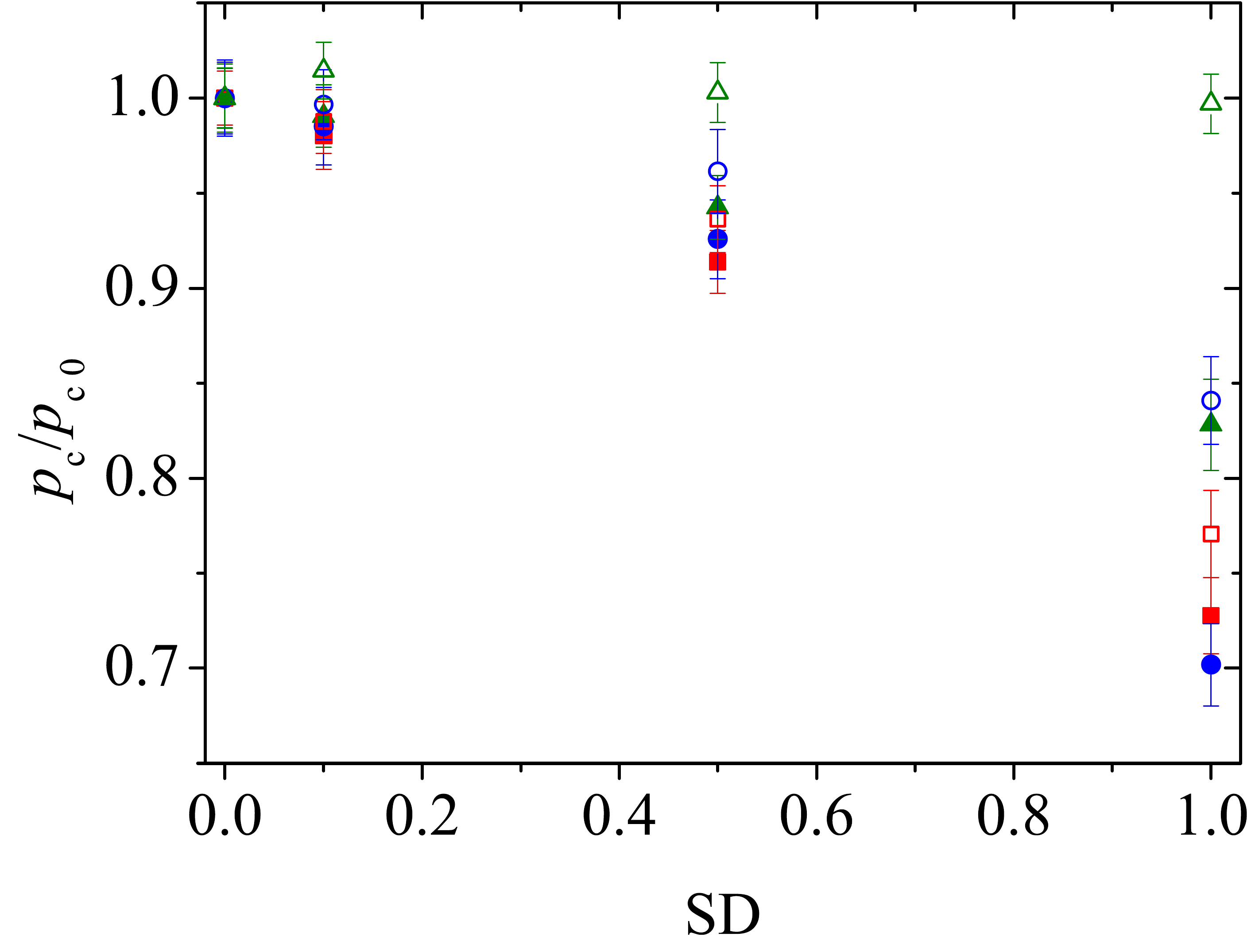}
\caption{\label{fig:pcm512}Examples of the normalized percolation threshold, $p_c/p_{c0}$, versus $\mathrm{SD}$ for three values of the standard deviation $\mathrm{SD}$ at $s$ = 0.5 (\opensquare, \fullsquare), 0.8 (\opencircle, \fullcircle), 1.0 (\opentriangle, \fulltriangle) (closed and open symbols correspond to the direction along alignment of the rods and in the perpendicular direction, correspondingly) and $m$ = 256. Here, $p_{c0}$ corresponds to $\mathrm{SD}$ = 0.}
\end{minipage}\hfill%
\begin{minipage}{0.45\textwidth}
\includegraphics[width=\textwidth]{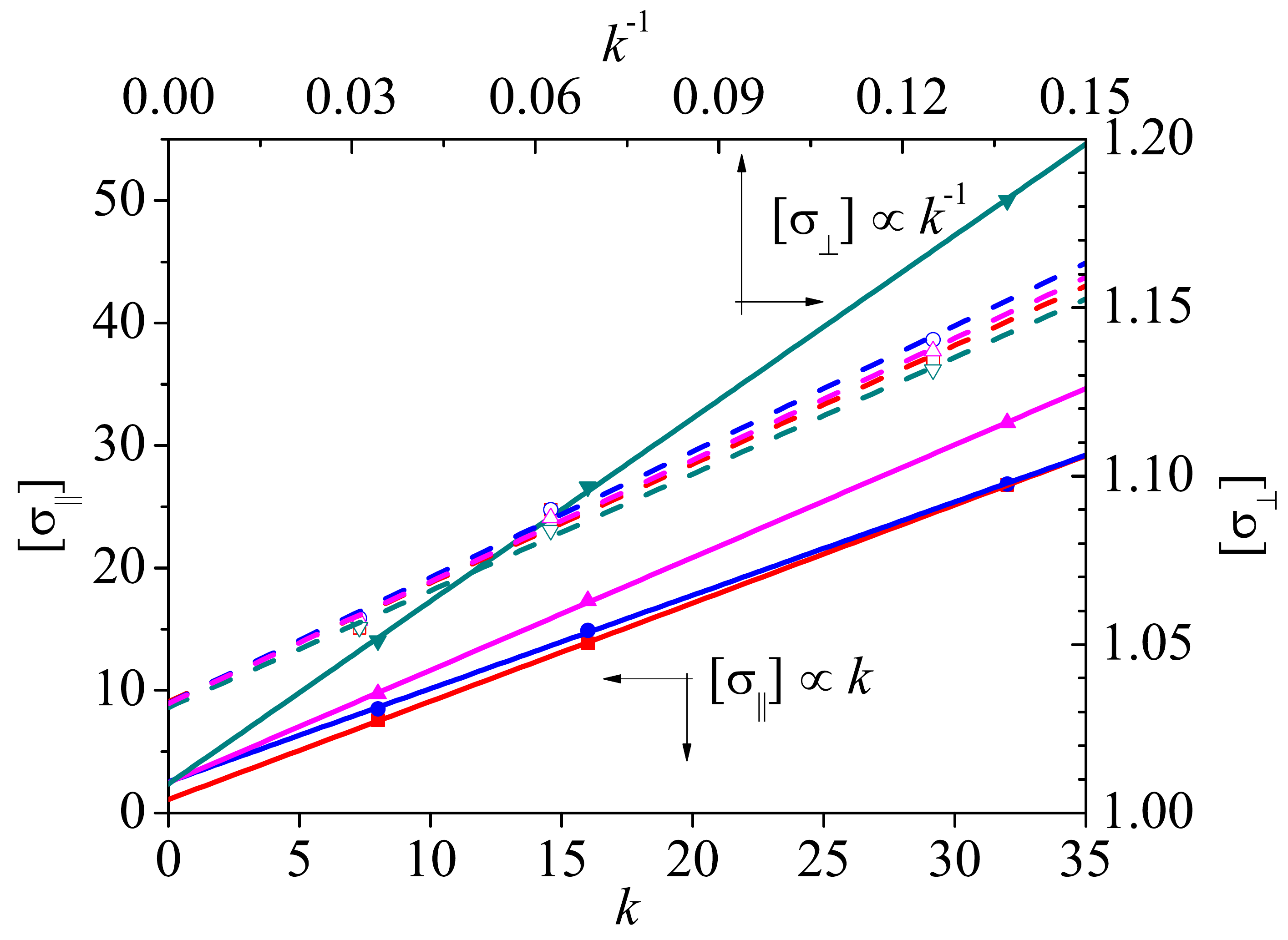}
\caption{\label{fig:IC}The intrinsic conductivity along the direction of alignment of the rods, $[\sigma_\parallel]$, vs the aspect ratio, $k$, of the rods (closed symbols, solid lines)  and the intrinsic conductivity in the perpendicular direction, $[\sigma_\perp]$  vs $1/k$ (open symbols, dotted lines) at $\mathrm{SD}$=0 (\opensquare, \fullsquare), 0.1 (\opencircle, \fullcircle), 0.5 (\opentriangle, \fulltriangle), 1 (\opentriangledown, \fulltriangledown) for the case of completely aligned rods ($s = 1$).}
\end{minipage}
\end{figure}

The results obtained for the values of the percolation threshold, $p_c$, for different values of the length dispersity and the order parameter ($s$ = 0.5, 0.8, 1.0) are summarized in~\fref{fig:pcm512}. Here the value of the percolation threshold is normalized to ones at $\mathrm{SD}=0$, the results are given both for along the direction of alignment of the rods (closed symbols) and in the perpendicular direction (open symbols) and $m=512$. The previously obtained conclusions are retained for all the investigated parameters $s$ and $\mathrm{SD}$, except for the case when all the rods are aligned in one direction ($s=1$) and the percolation threshold is determined in the direction perpendicular to the alignment direction. In this case, the value of the percolation threshold does not depend on the dispersity of the length of the rods, since the connectivity of the system in the direction perpendicular to the alignment does not depend on the length of the rods.

One of the important characteristics of the system is intrinsic conductivity; the results of its calculation are shown in~\fref{fig:IC}.
Here the intrinsic conductivity along the direction of alignment of the rods, $[\sigma_\parallel]$,  versus the aspect ratio of the rods, $k$, and the intrinsic conductivity in the perpendicular direction, $[\sigma_\perp]$  versus $1/k$  for the case of completely aligned rods ($s = 1$) are given at $\mathrm{SD}$=0, 0.1, 0.5, 1. The aspect ratio was calculated as $m/L$ for three values $m$=256, 512, 1024. The lines  are a linear approximation.

\begin{figure}[!htbp]
($a$)\includegraphics[width=0.45\textwidth]{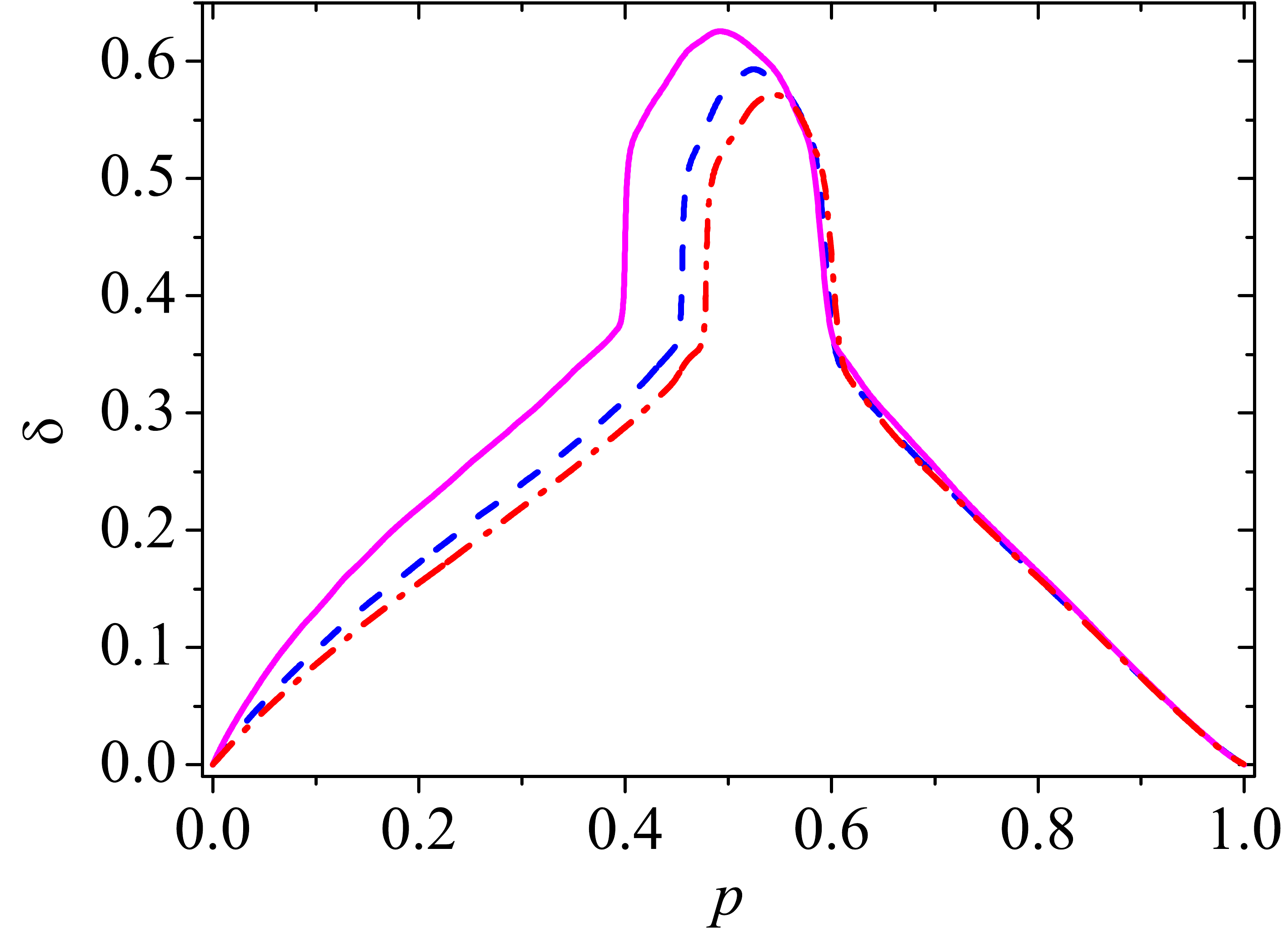}\hfill
($b$)\includegraphics[width=0.45\textwidth]{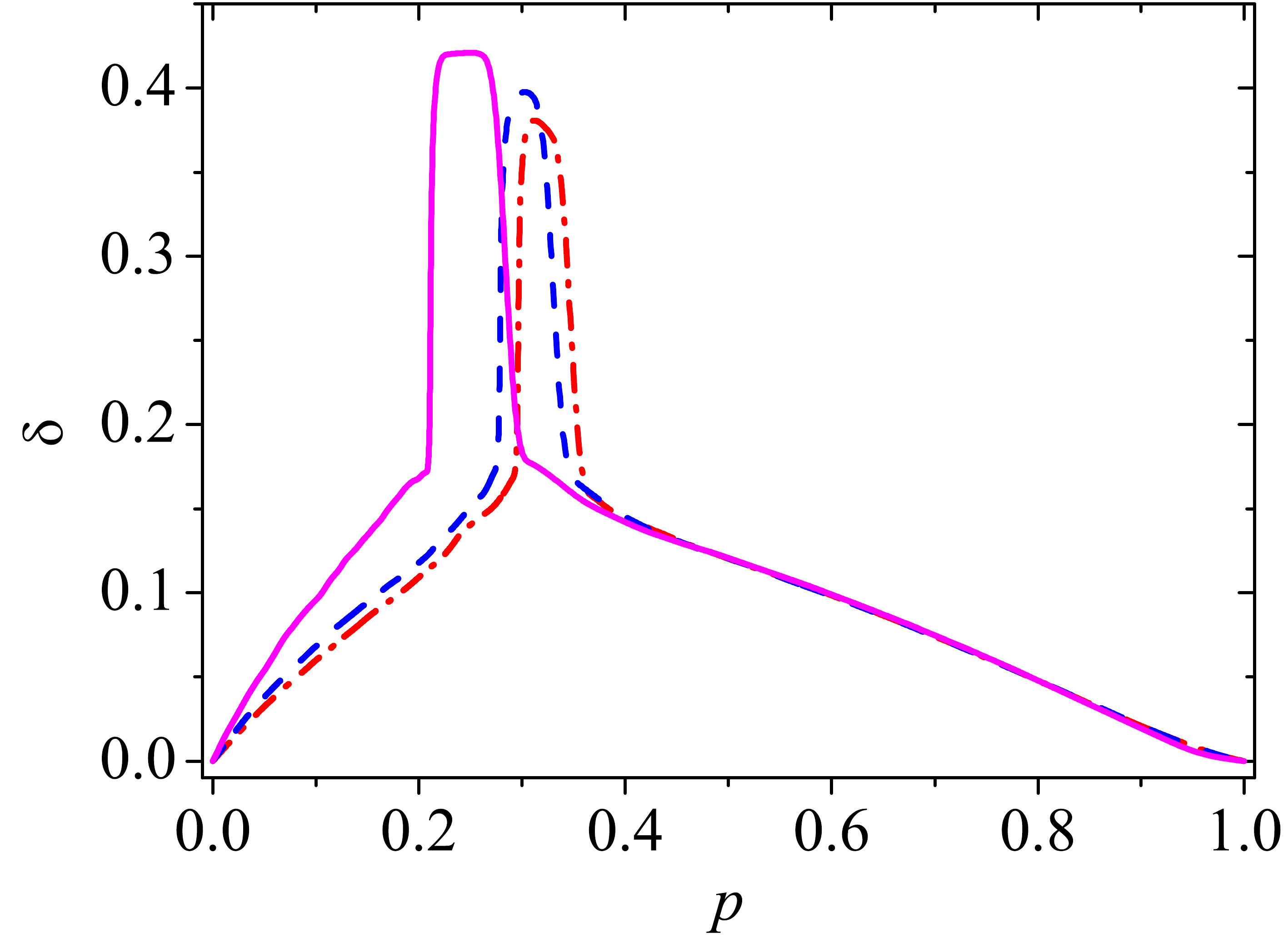}
\caption{\label{fig:deltam512}Example of the dependencies of the electrical anisotropy, $\delta$, vs the concentration, $p$, for different values of the length dispersity $\mathrm{SD} = 0.1 \chain, 0.5 \dashed, 1 \full$, $m = 512$ and $L = 32$: ($a$)~completely aligned rods $s = 1$; ($b$)~$s = 0.8$.}
\end{figure}

\Fref{fig:deltam512} presents examples of the electrical anisotropy ratio, $\delta$, versus the concentration, $p$, for $\mathrm{SD}=0.1, 0.5, 1$, $m=512$: ($a$) $s=1$, ($b$) $s=0.8$. In the case $s=0.8$, arrangement of the rods (\fref{fig:deltam512}($b$)) the maximum of the electrical anisotropy ratio is shifted to the region of smaller values $p$ and the interval of $p$ for large electrical anisotropy is smaller compared with completely aligned fillers ($s=1$) (\fref{fig:deltam512}($a$)).

\section{Conclusion}

Our computer simulation suggests that the dispersity of the filler length had significant impact only at the electrical conductivity of films along the direction of filler alignment; the electrical conductivity in the direction perpendicular to the filler alignment was insensitive to the dispersity of filler length. This dispersity had a positive impact on the figure of merit of films. It decreases the percolation threshold and, as a result, the critical concentration of fillers when the insulator--conductor phase transition occurs. In the vicinity of the phase transition, the electrical conductivity of the film exhibits high anisotropy.

\ack We acknowledge the funding from the Ministry of Education and Science of the Russian Federation, Project No.~3.959.2017/4.6( Y.Y.T., A.V.E., I.V.V., P.G.S., V.V.C.) and from the National Academy of Sciences of Ukraine, Projects No.~2.16.1.6. (0117U004046), 43/19-H, and 15F (0117U006352)  (N.I.L.).

\section*{References}
\bibliographystyle{iopart-num}
\bibliography{conductivity}

\end{document}